\documentclass[journal]{IEEEtran}
\usepackage{graphicx,dblfloatfix}
\usepackage{diagbox}
\usepackage{multirow}
\usepackage{amsmath,amssymb,amsfonts}
\usepackage{algorithmic}
\usepackage{threeparttable}
\usepackage{cite}
\usepackage{soul}
\usepackage{xcolor}
\begin{document}

\title{Automatic Segmentation and Visualization of Choroid in OCT with Knowledge Infused Deep Learning}

\author{Huihong Zhang, Jianlong Yang$^{*}$, Kang Zhou, Fei Li, Yan Hu, Yitian Zhao, Ce Zheng, Xiulan Zhang, and Jiang Liu
\thanks{This work was supported by Ningbo “2025 S\&T Megaprojects” (2019B10033), Zhejiang Provincial Natural Science Foundation (LQ19H180001), and Ningbo Public Welfare Science and Technology Project (2018C50049).}
\thanks{H. Zhang, J. Yang, L. Fang, Y. Zhao are with Cixi Institute of Biomedical Engineering, Ningbo Institute of Materials Technology and Engineering, Chinese Academy of Sciences, China. }
\thanks{F. Li and X. Zhang are with State Key Laboratory of Ophthalmology, Zhongshan Ophthalmic Center, Sun Yat-sen University, China.}
\thanks{K. Zhou is with School of Information Science and Technology, ShanghaiTech University, China.}
\thanks{C. Zheng is with Department of Ophthalmology, Xinhua Hospital Affiliated to Shanghai Jiao Tong University School of Medicine, Shanghai, China.}
\thanks{Y. Hu and J. Liu are with Department of Computer Science and Engineering, Southern University of Science and Technology, China.}
\thanks{H. Zhang is also with University of Chinese Academy of Sciences, China.}
\thanks{$^{*}$Corresponding author: Jianlong Yang (Email: yangjianlong@nimte.ac.cn)}
}

\maketitle

\begin{abstract}

The choroid provides oxygen and nourishment to the outer retina thus is related to the pathology of various ocular diseases. Optical coherence tomography (OCT) is advantageous in visualizing and quantifying the choroid \textit{in vivo}, because it does not suffer from the information contamination of the inner retina in fundus photography and scanning laser ophthalmoscopy, and the resolution deficiency in ocular ultrasound. However, its application in the study of the choroid is still limited for two reasons. (1) The lower boundary of the choroid (choroid-sclera interface) in OCT is fuzzy, which makes the automatic segmentation difficult and inaccurate. (2) The visualization of the choroid is hindered by the vessel shadows from the superficial layers of the inner retina. In this paper, we propose to incorporate medical and imaging prior knowledge with deep learning to address these two problems. We propose a biomarker infused global-to-local network, for the choroid segmentation. It leverages the thickness of the choroid layer, which is a primary biomarker in clinic, as a constraint to improve the segmentation accuracy. We also design a global-to-local strategy in the choroid segmentation: a global module is used to segment all the retinal and choroidal layers simultaneously for suppressing overfitting and providing global structure information, then a local module is used to refine the segmentation with the biomarker infusion. For eliminating the retinal vessel shadows, we propose a deep learning pipeline, which firstly use anatomical and OCT imaging knowledge to locate the shadows using their projection on the retinal pigment epthelium layer, then the contents of the choroidal vasculature at the shadow locations are predicted with an edge-to-texture two-stage generative adversarial inpainting network. The experiments shows the proposed method outperforms the existing methods on both the segmentation and shadow elimination tasks. We further apply the proposed method in a clinical prospective study for understanding the pathology of glaucoma, which demonstrates its capacity in detecting the structure and vascular changes of the choroid related to the elevation of intra-ocular pressure.
\end{abstract}


%
\IEEEpeerreviewmaketitle

\section{Introduction}
\indent The choriod, lying between the retina and the sclera, is the vascular layer which provides oxygen and nourishment to the outer retina \cite{atchison2000optics}. Because traditional imaging modalities like fundus photography and scanning laser ophthalmoscopy acquire 2D overlapping information of the retina and the choroid, the pathological changes of the choroid could not be precisely retrieved and evaluated. On the other hand, ocular ultrasound is able to do 3D imaging, but it needs to touch the eye and has a low spatial resolution.\\
\indent Optical coherence tomography (OCT) is a high-resolution non-invasive 3D imaging modality that could precisely separate the information of the underlying choroid from the inner retina, thus has been becoming a powerful tool to understand the role of the choroid in various ocular diseases \cite{laviers2014enhanced}. It has been shown that the thickness of the choroid layer extracted from OCT, is directly related to the incidence and severity of  predominate ocular diseases, such as pathological myopia \cite{wang2015choroidal}, diabetic retinopathy (DR) \cite{regatieri2012choroidal}, age-related macular degeneration (AMD) \cite{yiu2015relationship}, and glaucoma \cite{chen2014changes}.\\
\begin{figure}[b!]
    \centering
    \includegraphics[width=8cm]{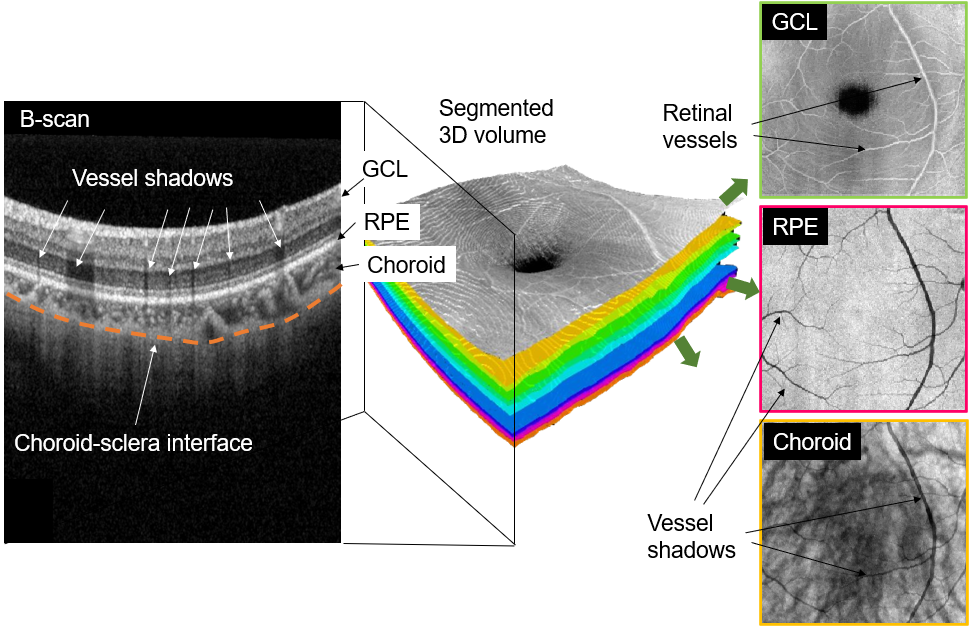}
    \caption{Demonstration of the choroid-sclera interface (orange dashed line) and the retinal vessels and their projection (arrows) on the layers including the GCL (green box), RPE (pink box) and choroidal vasculature (orange box). GCL: ganglion cell layer. RPE: retinal pigment epithelium.}
    \label{intro}
\end{figure} 
\indent Besides, the choroidal vasculature has also been applied in the study and diagnosis of ocular diseases. Agrawal \textit{et al.} found the choroidal vascularity index, which is extracted from binarized OCT B-scan, is related to the vascular status of DR \cite{tan2016choroidal}. The choroidal vessel density (CVD), extracted from binarized \textit{en face} choroid image, has been used in the evaluation of AMD \cite{zheng16} and central serous chorioretinopathy \cite{kuroda2016increased}. Wang \textit{et al.} further introduced the choroidal vascular volume, which combines the CVD and the choroidal thickness, is more sensitive in detecting proliferative DR \cite{wang2017diabetic}.\\
\indent However, the application of the choroidal biomarkers in clinic is still quite limited, which may be attributed to two primary reasons.  (1) The lower boundary of the choroid (choroid-sclera interface, CSI) in OCT is fuzzy, which makes the automatic segmentation difficult and inaccurate. (2) The visualization of the choroid is contaminated by the vessel shadows from the superficial layers of the inner retina. \\
\indent Figure~\ref{intro} is a demonstration of the CSI and the retinal vessels and their projection on the RPE and choroid layers. The position above the orange dashed line shows the fuzzy CSI in a B-scan. The anisotropy of the red blood cells inside the vessels cause strong forward attenuation of the probe light,thus bringing shadow-like dark tails to the underneath layers extending to the choroid and the sclera (white arrow). The center part of Fig. \ref{intro} is a segmented OCT volume, which could further be used to generate the en face images of each layer in the right side. The ganglion cell layer (GCL) possesses the retinal vessels (black arrows) and has high light reflectance (green box). The depth-projected vessel shadows (black arrows) turn dark on the vessel-absent retinal pigment epithelium (RPE) layer (pink box) and the choroid layer (orange box). It is evident that the shadows bring difficulties to the extraction of the choroidal vasculature.\\
\indent Due to their clinical significance, the automatic segmentation and visualization of the choroid have drawn numerous research interests recently \cite{hu2013semiautomated,tian2013automatic,alonso2013automatic,chen2015automated,sui2017choroid,masood2019automatic,cheng2019choroid}. However, the majority of the choroid segmentation methods are based on graph search \cite{li2005optimal}, which is restricted by the choice of a suitable graph-edge weight model \cite{sui2017choroid}. The inferior choice of the edge weight or the variation of OCT image features would cause inaccuracy in the choroid segmentation \cite{lee2013comparative}, so tedious manual inspection and correction are still required for clinical usage \cite{li2019upside}. The existing methods for eliminating the vessel shadows are based on the compensation of vessel-induced light attenuation \cite{girard2011shadow,mao2019deep}, but the effectiveness of this kind of A-line based method is limited to small vessels and capillaries in OCT retinal imaging. The large vessel shadows still have residue on the choroid \cite{yang2019enhancing}.\\
\indent To address these two problems, and inspired by the recent success of deep learning in medical image processing \cite{shen2017deep,litjens2017survey,gong2020statistical,dong2019land,kordestani2019failure}, we propose an automatic segmentation and visualization method for the choroid in OCT via knowledge infused deep learning. The main contributions of our work include:
\begin{itemize}
\item We propose a biomarker-infused global-to-local network (Bio-Net) for the choroid segmentation, which not only regularizes the segmentation via predicted choroid thickness, but also leverages a global-to-local segmentation strategy to provide global structure information and suppress overfitting. Their effectiveness has been validated via a comprehensive ablation study in Section \uppercase\expandafter{\romannumeral5}. 
\item For eliminating the retinal vessel shadows, we propose a deep-learning pipeline, which firstly locate the shadows using their projection on the retinal pigment epthelium layer, then the contents of the choroidal vasculature at the shadow locations are predicted with an edge-to-texture generative adversarial inpainting network.
\item The experiments shows the proposed method outperforms the existing methods on both the choroid segmentation and shadow elimination tasks.
\item We further apply the proposed method in a clinical prospective study for understanding the pathology of glaucoma, which demonstrates its capacity in detecting the structure and vascular changes of the choroid related to the elevation of intra-ocular pressure (IOP).
\end{itemize}
\indent The remainders of the paper are organized as follows. We review the existing techniques related to the proposed method in Section \uppercase\expandafter{\romannumeral2}. The methodology of the proposed method is presented in Section \uppercase\expandafter{\romannumeral3}. To validate the effectiveness and clinical significance of the proposed method, we conduct extensive experiments in Section \uppercase\expandafter{\romannumeral4}. We analyse and discuss the details of the proposed method in Section \uppercase\expandafter{\romannumeral5}, and draw our conclusion in Section \uppercase\expandafter{\romannumeral6}.
\section{Related works}
\begin{figure*}[h!]
    \centering
    \includegraphics[width=16cm]{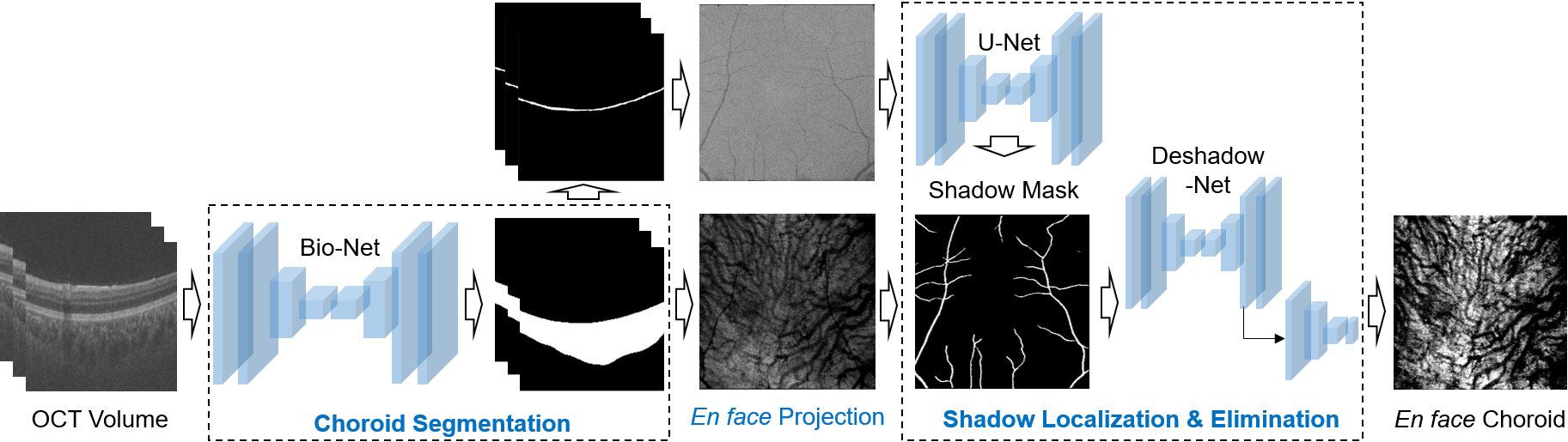}
    \caption{Illustration of the framework of our method, which primarily includes the choroid segmentation using the proposed Bio-Net, \textit{en face} projection, and the shadow localization and elimination using the U-Net for shadow location mask generation and the Deshadow-Net for shadow elimination. The choroid layer of a OCT volume is firstly segmented by the Bio-Net. We can get the RPE layer by moving the upper boundary of the choroid 20 $\mu$m upward. Then the OCT volume is projected into the 2D \textit{en face} plane with the mean value projection along the axial direction, for generating the \textit{en face} RPE and choroid images. The vessel shadows in the RPE projection image is segmented with the U-Net to locate their positions. Finally, the shadow location mask in combination with the \textit{en face} choroid image are inputted into the Deshadow-Net for the shadow elimination.}
    \label{framework}
\end{figure*}
\subsubsection{Automatic Choroid Segmentation} 
The segmentation of the retinal layers in OCT has been explored since the commercialization of spectral domain (SD) OCT \cite{garvin2009automated,
lee2009segmentation,yang2010automated,zhang2012automated,zhang2015validity,gerendas2014three,philip2016choroidal}, but the segmentation of the choroid layer was usually not included in these studies, which may be attributed to the fussy CSI (as a comparison, the inner retinal layers usually have sharp and smooth boundary as shown in Fig.~\ref{intro}). Hu \textit{et al.} adapted the graph search algorithm to semi-automatically identify the choroidal layer in SD-OCT \cite{hu2013semiautomated}. Tian \textit{et al.} segmented the CSI by finding the shortest path of the graph formed by valley pixels using Dijkstra’s algorithm \cite{tian2013automatic}. Alonso \textit{et al.} developed an algorithm that detected the CSI based on OCT image enhancement and a dual brightness probability gradient \cite{alonso2013automatic}. Chen \textit{et al.} generated a gradual intensity distance image. Then an improved 2D graph search method with curve smooth constraints was used to obtain the CSI segmentation \cite{chen2015automated}. Wang \textit{et al.} segment the choroid layer using a Markov random field and distance regularization and edge constraint terms embedded level set method \cite{wang2017}.\\
\indent With the development of deep learning techniques, many classical models for classification and segmentation tasks were proposed, such as CNN \cite{krizhevsky2012imagenet}, FCN \cite{long2015fully}, SegNet \cite{badrinarayanan2017segnet}, and U-Net \cite{ronneberger2015u}.
These classical deep learning networks or their 3D versions were directly used to segment the choroid \cite{kugelman2019automatic,masood2019automatic,shah2017simultaneous,cciccek20163d}, or utilized as a feature extractor and combined with other methods\cite{sui2017choroid,chen2017automated}. Sui \textit{et al.} combined the graph search with convolutional neural network (CNN) by using the used CNN to decide the edge weights in the graph search \cite{sui2017choroid}. Masood \textit{et al.} converted the segmentation tasks into a binary classification task, which extracted the choroid part of OCT images into patches with or without the CSI \cite{masood2019automatic}. The U-Net may be the most successful architecture for medical image segmentation to date \cite{ronneberger2015u}. Cheng \textit{et al.} proposed an improved U-Net with refinement residual block and channel attention block for the choroid segmentation \cite{cheng2019choroid}.\\
\subsubsection{Vessel Shadow Removal in OCT} In 2011, Girard \textit{et al.} developed an attenuation compensation (AC) algorithm to remove the OCT vessel shadows and enhance the contrast of optic nerve head \cite{girard2011shadow}. This algorithm was then employed in the calculation of the attenuation coefficients of retinal tissue \cite{vermeer2014depth}, enhancing the visibility of lamina cribrosa \cite{mari2013enhancement}, and improving the contrast of the choroid vasculature and the visibility of the sclera-choroid interface \cite{zhou2018attenuation,vupparaboina2018quantitative}.\\
\indent Very recently, Mao \textit{et al.} analysed the energy profile in each A-line and automatically compensated the pixel intensity of locations underneath the detected blood vessel \cite{mao2019deep}. However, both of these methods perform well for the removal of small vessel shadows but unable to handle the large vessel shadows, which would lead to shadow residue on the choroid \cite{yang2019enhancing}.\\
\subsubsection{Inpainting/Object Removal} After locating the vessel shadows, we propose to use inpainting techniques, which is also referred as object removal. Here the object to be removed is the vessel shadows. Inpainting techniques have been extensively studied and applied in various computer vision and pattern recognition related fields (see \cite{guillemot2013image,elharrouss2019image} and the references therein). Early inpainting techniques primarily filled the targeted area with information from similar or closest image parts, such as exemplar-based inpainting (EBI) \cite{criminisi2004region}, or used higher-order partial differential equations to propagate the information of surrounding areas into the targeted area, such as coherence transport inpainting (CTI) \cite{bornemann2007fast}.\\
\indent Deep learning, especially generative adversarial network (GAN) \cite{goodfellow2014generative}, is a powerful tool for image synthesis \cite{zhu2017unpaired}, which has also shown its superiority in image inpainting \cite{yeh2017semantic,yu2018generative,nazeri2019edgeconnect}. Yeh \textit{et al.} used a GAN model to search for the closest encoding of the corrupted image in the latent image manifold using context and prior losses, then passed the encoding through the GAN model to infer the missing content \cite{yeh2017semantic}. Yu \textit{et al.} utilized contextual attention on surrounding image features as references during GAN training to make better predictions \cite{yu2018generative}. Nazeri \textit{et al.} proposed a two-stage GAN inpainting method, which comprises of an edge generator followed by an image completion network. The edge generator hallucinates edges of the missing region of the image, and the image completion network fills in the missing regions using hallucinated edges as a priori \cite{nazeri2019edgeconnect}.

\section{Methodology}
Figure~\ref{framework} is an illustration of the framework of our method, which primarily includes the choroid segmentation using the proposed Bio-Net, \textit{en face} projection, and the proposed shadow localization and elimination pipeline. We use the U-Net for shadow location mask generation and the Deshadow-Net for shadow elimination. The Deshadow-Net follows the architecture of the two-stage inpainting GAN in \cite{nazeri2019edgeconnect}. The choroid layer of a OCT volume is firstly segmented by the Bio-Net. We can get the RPE layer by moving the upper boundary of the choroid 20 $\mu$m upward. Then the segmented RPE and choroid regions are projected into the 2D \textit{en face} plane with the mean value projection along the axial direction, for generating the \textit{en face} RPE and choroid projection images. The vessel shadows in the RPE projection image is segmented with the U-Net to locate their positions. Finally, the shadow location mask in combination with the \textit{en face} choroid image are inputted into the Deshadow-Net for the shadow elimination.
\begin{figure}[!h]
    \centering
    \includegraphics[width=\linewidth]{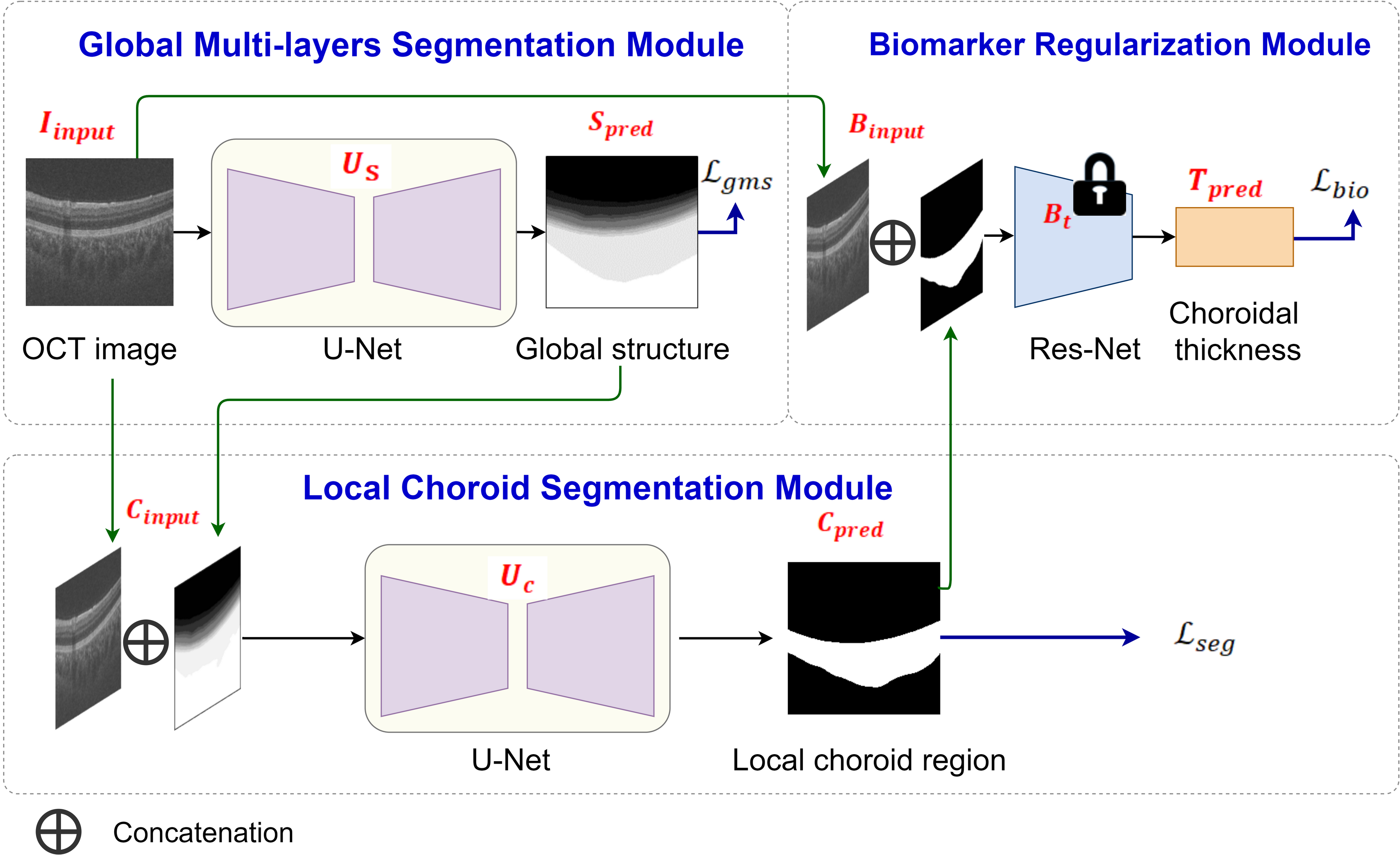}
    \caption{Illustration of the Bio-Net. It includes three modules. A global multi-layers segmentation module (GMS) is employed to segment all the layers (both retinal and choroidal layers) in OCT image $I_{input}$. The output of the GMS $S_{pred}$, which can be regarded as the global structure of $I_{input}$, is concatenated with $I_{input}$ and inputted into a local choroid segmentation module (LCS). Different from the GMS, the LCS only segments the choroid region (from the Bruch's membrane to choroid-sclera interface), which is the output of the Bio-Net. On the other hand, a biomarker regularization module (Bio) is used to regularize the segmentation of the choroid region $C_{pred}$. The $I_{input}$ is concatenated as the input of the Bio for providing texture information.
}
 \label{bionet}
\end{figure}

\subsection{Bio-Net for Choroid Segmentation}
 
Figure~\ref{bionet} is an illustration of the Bio-Net. It includes three modules. A global multi-layers segmentation module (GMS) is employed to segment all the layers (both retinal and choroidal layers) in OCT image $I_{input}$. The output of the GMS $S_{pred}$, which can be regarded as the global structure of $I_{input}$, is concatenated with $I_{input}$ and inputted into a local choroid segmentation module (LCS). Different from the GMS, the LCS only segments the choroid region (from the Bruch's membrane to choroid-sclera interface), which is the output of the Bio-Net. On the other hand, a biomarker regularization module (Bio) is used to regularize the segmentation of the choroid region $C_{pred}$. Also, the $I_{input}$ is concatenated as the input of the Bio for providing texture information. The details of these modules are given below.
 \subsubsection{Global Multi-Layers Segmentation Module} 
The GMS is to segment retinal and choroidal layers following their anatomical characteristics in OCT images. It is mainly used to obtain the global structure information, which could be an effective auxiliary to improve local segmentation accuracy \cite{chen2019collaborative}. Besides, as a multi-tasking network, the segmentation of different layers are constrained with each other, which can reduce over-fitting and improve robustness \cite{ruder2017overview}.\\
\indent The GMS takes the $I_{input}$ as input and generates the multi-layer segmentation result $S_{pred}$ via a deep neural network $\textbf{U}_{s}$, $S_{pred}=\textbf{U}_{s}(I_{input})$. The architecture of $\textbf{U}_{s}$ is the U-Net \cite{ronneberger2015u}. The segmentation is optimized via a cross entropy loss $\mathcal{L}_{gms}$, which could calculated as:
\begin{equation}   
   \begin{aligned}
 \mathcal{L}_{gms}=-\sum_{i}S_{gt}ln(S_{pred}),
 \end{aligned}
\end{equation}
where $S_{gt}$ denotes the manually-annotated ground truth of all retinal and choroidal layers. $i$ denotes the index of layers. 
\subsubsection{Local Choroid Segmentation Module}
The LCS takes the concatenation of the global segmented structure $S_{pred}$ from the GMS and the original OCT image $I_{input}$ as input. It segments the choroid region $C_{pred}$ via another segmentation network $\textbf{U}_c$, $C_{pred}=\textbf{U}_c(C_{input})$. $C_{gt}$ is the manually-annotated ground truth of the choroid region, which contains two categories of pixels (choroid and background). The $\textbf{U}_c$ also uses the U-Net architecture\cite{ronneberger2015u} and a binary cross entropy loss $\mathcal{L}_{seg}$ for optimization:
\begin{equation}
    \mathcal{L}_{seg} = -[C_{gt}ln(C_{pred}) + (1-C_{gt})ln(1-C_{pred})].
\end{equation}
\indent In addition to the $\mathcal{L}_{seg}$, the LCS is also regularized by the Bio module and its loss $\mathcal{L}_{bio}$. So the total loss of the LCS could be written as:
\begin{equation}
	\mathcal{L}_{lcs} = \lambda_{seg}\mathcal{L}_{seg}+\lambda_{bio}\mathcal{L}_{bio},
\end{equation}
where $\lambda_{seg}$, $\lambda_{bio}$ denote the corresponding hyper-parameters. 
\subsubsection{Biomarker Regularization Module}
The Bio further uses the thickness of choroid layer to regularize the segmentation of the Bio-Net. It predicts a thickness vector $T_{pred}\in  \mathbb R ^{W\times 1}$, where $W$ is the width of OCT images and also refers to the number of A-lines. The Bio takes the concatenation of the original OCT image $I_{input}$ and the predicted choroid region $C_{pred}$ as input $B_{input}$. So $T_{pred}=\textbf{B}_{t}(B_{input})$, where $\textbf{B}_{t}$ is a biomarker regression network and uses the architecture of the ResNet-18 \cite{he2016deep}. Its optimization is to minimize the $\mathcal{L}1$ distance between the $T_{pred}$ and the thickness ground truth $T_{gt}$:
\begin{equation}
\mathcal{L}_{bio}=\lVert T_{pred}-T_{gt}\rVert_1.
\end{equation}
The $T_{gt}$ is generated from the $C_{gt}$ by counting the pixel number of the choroid region along each A-line.\\
\\
\indent So, the total loss $\mathcal{L}_{total}$ of the proposed Bio-Net is:
\begin{equation}
\begin{aligned}
 \mathcal{L}_{total} =&\lambda_{gms}\mathcal{L}_{gms}+ \mathcal{L}_{lcs}\\
						=&\lambda_{gms}\mathcal{L}_{gms}+ \lambda_{seg}\mathcal{L}_{seg} + \lambda_{bio}\mathcal{L}_{bio},
\end{aligned}
\end{equation}
where $\lambda_{gms}$ denotes the hyper-parameter of the GMS.
\subsubsection{Training and testing procedures}
\begin{figure}[h!]
\centering
\includegraphics[width=\linewidth]{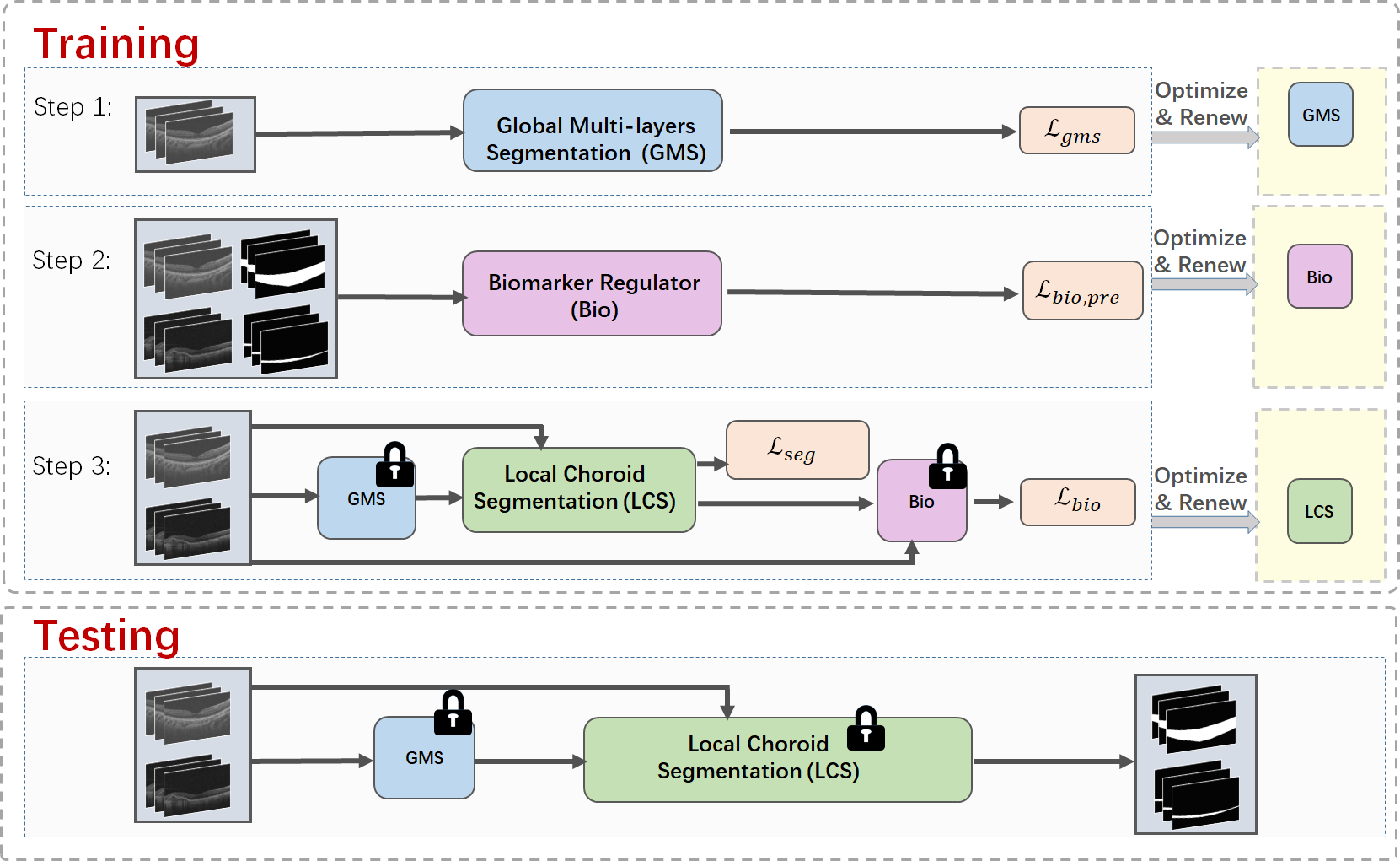}
\caption{
Training and testing procedures of the Bio-Net. We firstly pre-train the GMS and Bio modules. Then these trained modules are frozen in the training of the LCS. In testing, we only employ the trained GMS and LCS. The former is used generate the global structure and the latter is for achieving the choroid segmentation results.}
\label{steps}
\end{figure}
Figure~\ref{steps} illustrates the training and testing procedures of the Bio-Net. We firstly pre-train the GMS and Bio modules. Then these trained modules are frozen in the training of the LCS. Note that the pre-training of the Bio is different from its deployment in the regularization of the choroid segmentation, which is because the predicted choroid region $C_{pred}$ is not available at the stage of pre-training. Here we use pseudo labels of choroid region to generate the thickness ground truth $T_{gt}$, which is derived from a pre-trained U-Net. This U-Net is trained for the choroid segmented by only using $I_{input}$ and $C_{gt}$. Finally, the LCS in combination with the frozen GMS and Bio, is trained in an end-to-end manner.\\
\indent In the testing stage, we only employ the trained GMS and LCS. The former is used generate the global structure and the latter is for achieving the choroid segmentation result.
\subsection{Shadow Removal Pipeline}
\begin{figure*}[!h]
    \centering
    \includegraphics[width=16cm]{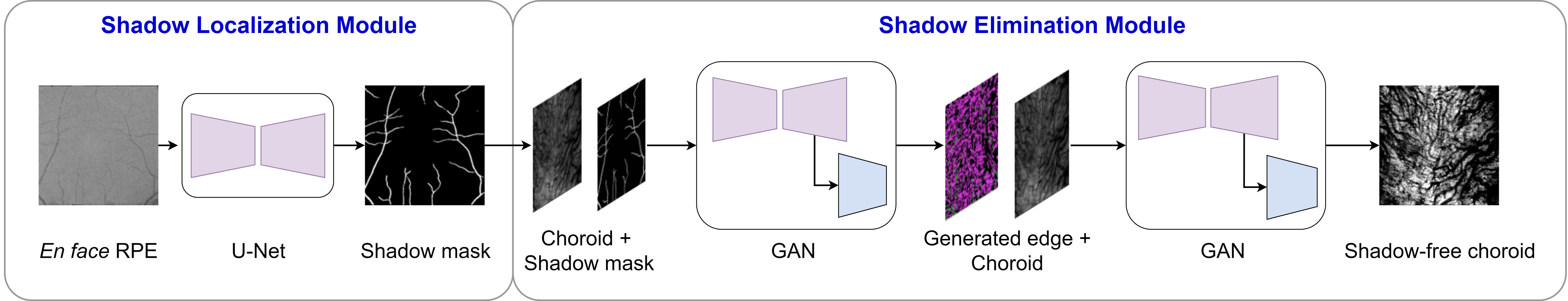}
    \caption{Illustration of the shadow elimination pipeline. We firstly segment the retinal vessel shadows from the \textit{en face RPE} image with the U-Net \cite{ronneberger2015u}. The generated shadow mask could be used to locate the shadows in a OCT volume. Then the shadow mask in combination with the \textit{en face} choroid image are fed into the shadow elimination module, namely the Deshadow-Net, to get a shadow-free choroid image.}
    \label{deshadow}
\end{figure*}

Different from the previous shadow elimination methods that could not eliminate the shadows from large vessel \cite{girard2011shadow,mao2019deep}, we propose a novel method that is able to remove the shadow without the limitation in vessel caliber. It firstly locates the vessel shadows then uses image inpainting techniques to repair the shadow-conterminated areas. As shown in Fig.~\ref{deshadow}, we segment the retinal vessel shadows from the \textit{en face RPE} image with the U-Net \cite{ronneberger2015u}. The generated shadow mask could be used to locate the shadows in a OCT volume. Then the shadow mask in combination with the \textit{en face} choroid image are fed into the shadow elimination module, namely the Deshadow-Net, to get a shadow-free choroid image.
\subsubsection{Shadow Localization}
\begin{figure}[!t]
    \centering
    \includegraphics[width=8cm]{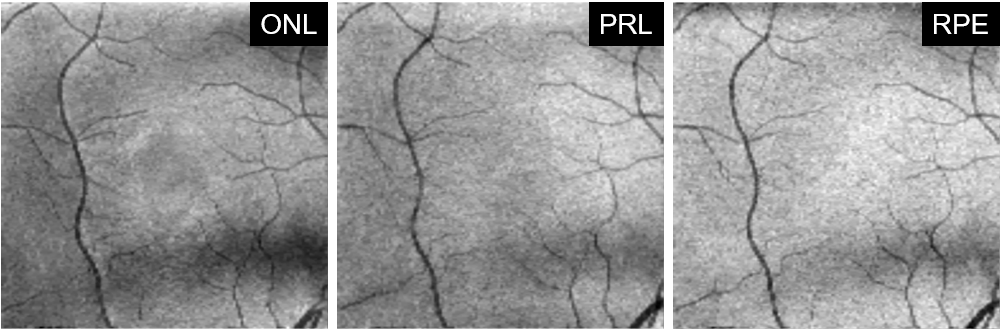}
    \caption{Comparison of the RPE layer with other avascular layers including outer nuclear layer (ONL) and photoreceptor layer (PRL) in OCT imaging.}
    \label{rpe}
\end{figure}
The idea of using RPE to locate the vessel shadows is inspired by two medical and imaging knowledge. (1) The retinal layers below the outer plexiform layer and above the BM are avascular \cite{campbell2017detailed}, so any vessel-like structure appears on these layers are the projected shadows. (2) As shown in Fig.~\ref{rpe}, the RPE layer has the highest OCT light reflectance and best shadow contrast compared with other avascular layers including outer nuclear layer (ONL) and photoreceptor layer (PRL).\\
\indent To fully locate the shadow mask on the \textit{en face} choroid, we further enhance the U-Net segmentation results with morphological manipulation including dilation and erosion. 
\subsubsection{Shadow Removal}
As demonstrated in Fig.~\ref{deshadow}, the Deshadow-Net is a cascade of two GANs. Each GAN has a pair of generator and discriminator. The generators follow the architecture in \cite{johnson2016perceptual} and the discriminators use a $70\times70$ PatchGAN architecture \cite{isola2017image}. The inputs of the Deshadow-Net are the shadow-contaminated choroid image and the shadow location mask. Before entering the first GAN, the structure feature of the \textit{en face} choroid is extracted with the Canny edge detector \cite{canny1986computational}. The first GAN is employed to generate the structure (edge) information in the shadow mask areas. The second GAN uses the edges of the choroidal vasculature generated in the first GAN as a prior, to fill the texture information of the choroid. Then we could get the shadow-free choroid from the output of the second GAN. 
\section{Experiments}
\subsection{Automatic Choroid Segmentation}
\subsubsection{Dataset}

Macular-centered 3D OCT volumes were collected  using the Topcon system from 20 different normal eye subjects. Each 3D volume contains 256 non-overlapping B-scans, covering $6\times 6 \times 2$ mm$^3$ region. Each B-scan has 512 A-lines with 992 pixels in each A-line.
Since the B-scans in the same volume have high similarity, Cheng \textit{et al.} \cite{cheng2016speckle} annotates the boundary information for $1/4$ of the B-scans, thus $256/4\times20=1280$ B-scans are used. 
The boundaries in each image were marked independently by two groups of trained professionals with the supervision of senior ophthalmologists. The $20$ volumes were randomly divided into a 12-volume training set, a 4-volume validation set, and a 4-volume test set. We also collected two diseased volumes from two subjects with different levels of choroidal neovascularization (CNV). One volume has 58 B-scans and was used in the training. The other volume has 21 B-scans and was used in the testing.
\indent Note that the data throughout this paper was collected from Topcon OCT systems, so we did not consider to remedy the domain discrepancy caused by manufacturers in the proposed method. For using it in the scenerios that the OCT systems are from different manufacturers, domain adaptation methods \cite{chai2019perceptual,mahmood2018unsupervised} have been used for achieving superior segmentation performance.
\subsubsection{Implementation}
\begin{figure*}[!]
    \centering
    \includegraphics[width=18cm]{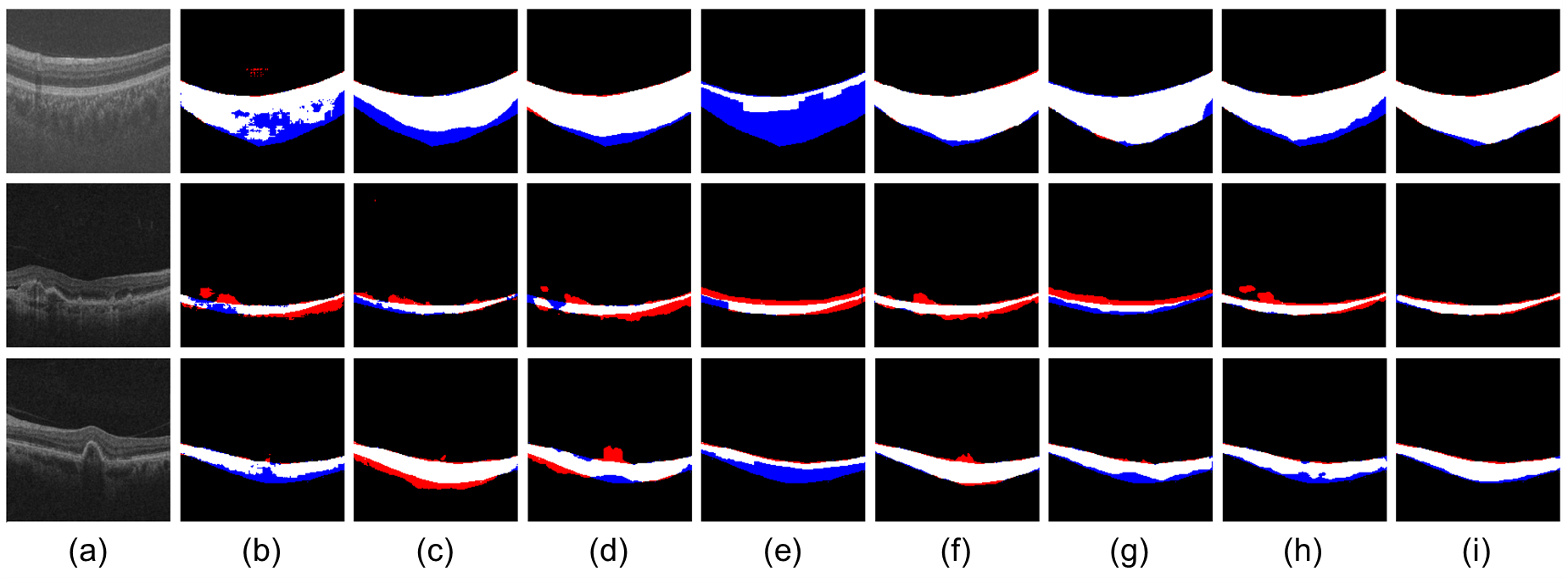}
    \caption{The visual examples of the choroid segmentation. From left to right: (a) is the input OCT B-scans, and from (b) to (i) are the segmentation results using the Long \textit{et al.} \cite{long2015fully}, Badrinarayanan \textit{et al.} \cite{badrinarayanan2017segnet}, Ronneberger \textit{et al.} \cite{ronneberger2015u}, Mazzaferri \textit{et al.}\cite{mazzaferri2017open}, Cheng \textit{et al.} \cite{cheng2019choroid},  Venhuizen \textit{et al.} \cite{Venhuizen2017}, Gu \textit{et al.} \cite{gu2019net}, and the proposed Bio-Net, respectively. White indicates the choroid region that is correctly segmented, red indicates excessive segmentation, and blue indicates insufficient segmentation.}
    \label{choroid1}
\end{figure*}
\begin{table*}[h!]
\centering
\caption{Quantitative Comparison of different choroid segmentation methods on normal samples.}\label{q1}
\begin{tabular}{llllll}
\hline
\multirow{2}{*}{Method}& \multicolumn{2}{c}{AUSDE (pixels)} & \multirow{2}{*}{TD (pixels)} & \multirow{2}{*}{Dice(\%)} & \multirow{2}{*}{IoU(\%)} \\ \cline{2-3}
& BM              & CSI              &                              &                           &                          \\ \hline
Long \textit{et al.} \cite{long2015fully}         & 1.07$\pm$0.01        & 8.02$\pm$0.10       & 7.91$\pm$0.11   & 87.55$\pm$0.14  &77.86$\pm$0.22   \\
Badrinarayanan \textit{et al.} \cite{badrinarayanan2017segnet}       & 2.49$\pm$0.30        & 7.77$\pm$0.38       & 6.27$\pm$0.11   & 88.57$\pm$0.17 &79.49$\pm$0.27   \\
Ronneberger \textit{et al.} \cite{ronneberger2015u}        & 1.45$\pm$0.02        & 5.87$\pm$0.04       & 5.48$\pm$0.06   & 90.26$\pm$0.08   &82.25$\pm$0.13   \\
Mazzaferri \textit{et al.}\cite{mazzaferri2017open}   & 37.82 $\pm$0.52      & 56.00 $\pm$0.65     & 24.64 $\pm$0.23 & 59.56  $\pm$0.38   &42.41  $\pm$0.38  \\
Cheng \textit{et al.} \cite{cheng2019choroid}      & 0.62$\pm$0.00        & 5.13$\pm$0.30       & 5.22$\pm$0.30   & 91.37$\pm$0.47    &84.12$\pm$0.79  \\
Venhuizen \textit{et al.} \cite{Venhuizen2017}  & 1.12$\pm$0.10        & 5.11$\pm$0.14       & 4.96$\pm$0.07   & 91.45$\pm$0.14  &84.25$\pm$0.24  \\
Gu  \textit{et al.} \cite{gu2019net}        & 0.88$\pm$0.06        & 4.61$\pm$0.07       & 4.73$\pm$0.07   & 92.03$\pm$0.10   &85.24$\pm$0.17\\ \hline
Proposed method     & 0.77$\pm$0.02        & 4.31$\pm$0.02       & 4.30$\pm$0.02   & 92.74$\pm$0.02  &86.47$\pm$0.04  \\ \hline
\end{tabular}
\end{table*}

\begin{table*}[b!]
\centering
\caption{Quantitative Comparison of different choroid segmentation methods on Samples with CNV.}\label{q3}
\begin{tabular}{llllllll}
\hline
\multirow{2}{*}{Method} & \multicolumn{2}{c}{AUSDE (pixels)} & \multirow{2}{*}{TD (pixels)} & \multirow{2}{*}{Dice(\%)} & \multirow{2}{*}{IoU(\%)} \\ \cline{2-3}
& BM              & CSI              &                              &                           &                          \\ \hline
Long \textit{et al.} \cite{long2015fully}                & 39.09$\pm$24.59    & 16.20$\pm$42.11    & 7.82$\pm$13.44                  & 74.45$\pm$14.22              & 61.00$\pm$15.20             \\
Badrinarayanan   \textit{et al.} \cite{badrinarayanan2017segnet}      & 30.70$\pm$23.21    & 34.67$\pm$25.50     & 4.50$\pm$2.31                   & 80.35$\pm$4.87               & 67.43$\pm$6.73              \\
Ronneberger \textit{et al.}  \cite{ronneberger2015u}        & 38.53$\pm$16.74    & 42.70$\pm$18.46     & 4.61$\pm$1.72                   & 82.57$\pm$16.17              & 72.73$\pm$17.58             \\
Mazzaferri \textit{et al.} \cite{mazzaferri2017open}       & 53.63$\pm$63.63    & 59.40$\pm$70.07      & 5.82$\pm$6.53                   & 84.68$\pm$25.04              & 79.07$\pm$26.98             \\
Cheng \textit{et al.}   \cite{cheng2019choroid}          & 46.16$\pm$12.80    & 50.70$\pm$14.02     & 4.90$\pm$1.29                   & 89.29$\pm$1.01               & 80.66$\pm$1.63              \\
Venhuizen \textit{et al.}    \cite{Venhuizen2017}       & 24.12$\pm$16.90    & 26.85$\pm$18.25     & 3.13$\pm$1.34                   & 88.86$\pm$0.55               & 79.96$\pm$0.89              \\
Gu  \textit{et al.}      \cite{gu2019net}             & 24.23$\pm$17.92    & 26.82$\pm$19.37     & 3.02$\pm$1.42                   & 89.59$\pm$0.63               & 81.15$\pm$1.03              \\ \hline
Proposed   method         & 16.59$\pm$12.58    & 18.33$\pm$13.53     & 2.07$\pm$0.99                   & 90.22$\pm$0.90               & 82.19$\pm$1.47              \\ \hline
\end{tabular}
\end{table*}
Our Bio-Net is implemented by using PyTorch library \cite{paszke2019pytorch} in the Ubuntu 16.04 operating system and the training was performed with NVIDIA GeForce GTX 1080 Ti GPU. We utilize horizontal flipping and rotation to augment the data. We resize all the B-scans to $192\times192$ pixels. The batch size is 4 and the optimizer is Adam \cite{kingma2014adam}. The initial value of the learning rate is 0.01, and then the learning rate is reduced to 1/10 of the original when the number of iterations is 40, 80, 160, and 240, respectively. 
We set the hyper-parameter $\lambda_{seg,multi-layers}=1$, $\lambda_{seg,choroid}=1$, $\lambda_{bio,choroid}=0.01$. It took about 4 hours for each training of the Bio-Net.
\subsubsection{Evaluation Metrics}
We employ dice index (Dice), intersection-over-union (IoU), average-unsigned-surface-detection-error (AUSDE), and thickness difference (TD) to quantitatively evaluate the performance of the Bio-Net. The Dice and IoU show the proportion of the overlap between the segmented choroid region and the ground truth (larger is better). The AUSDE \cite{xiang2018automatic} represents the pixel-wise mismatch between the segmented choroid boundary and the ground truth (smaller is better). We give the AUSDEs of the BM and CSI separately.
\subsubsection{Results}
\indent Figure~\ref{choroid1} shows the visual examples of the choroid segmentation. We compared the proposed method with the existing graph-search-based and deep-learning-based choroid segmentation methods. From left to right: (a) is the input OCT B-scans, and from (b) to (i) are the segmentation results using the Long \textit{et al.} \cite{long2015fully}, Badrinarayanan \textit{et al.} \cite{badrinarayanan2017segnet}, Ronneberger \textit{et al.} \cite{ronneberger2015u}, Mazzaferri \textit{et al.}\cite{mazzaferri2017open}, Cheng \textit{et al.} \cite{cheng2019choroid},  Venhuizen \textit{et al.} \cite{Venhuizen2017}, Gu \textit{et al.} \cite{gu2019net}, and the proposed Bio-Net, respectively. White indicates the choroid region that is correctly segmented, red indicates excessive segmentation, and blue indicates insufficient segmentation. As demonstrated in the figure, all of the methods have better performance in the segmentation of the BM than the CSI, which may be because the BM has bright and sharp features for recognition while the CSI is dark and fussy. Their quantitative comparison is listed in Table~\ref{q1}. Our Bio-Net outperforms other methods, which may suggest the the infusion of the biomarker prior and the global-to-local strategy contribute to the improvement of the segmentation.\\
\indent We further verify the performance superiority by using the diseased (CNV) data. The results are shown in Table~\ref{q3}. We can see the results using the proposed method are mostly better than others, although the metrics are worse than the results of normal cases. 
\subsection{Shadow Localization and Removal}
\begin{figure}[!h]
    \centering
    \includegraphics[width=8cm]{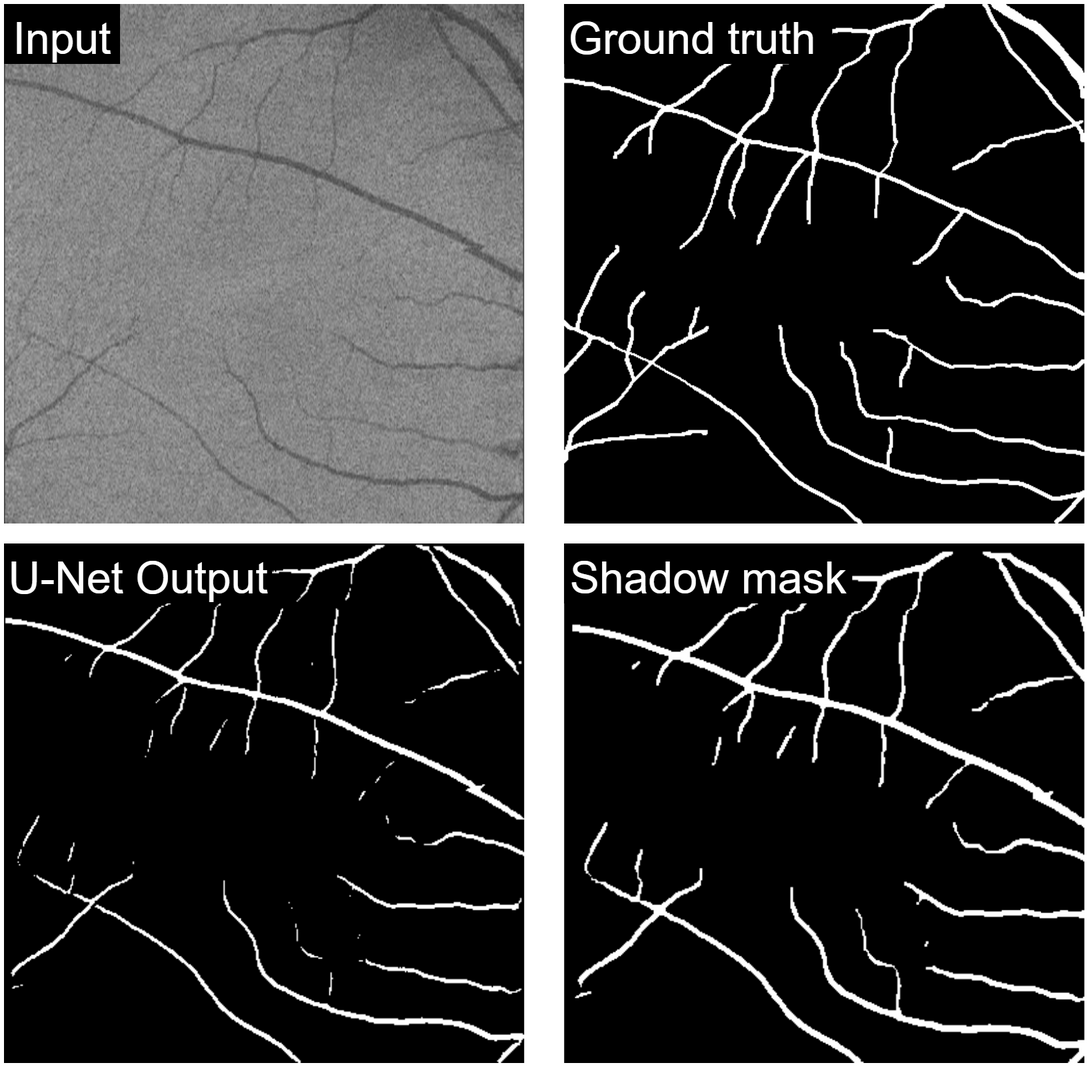}
    \caption{An example of using the U-Net in the segmentation of the retinal vessel shadows on the RPE projection image. The U-Net output is further processed to retrieve the shadow mask using dilation and erosion.}
    \label{shadow1}
\end{figure}
\begin{figure}[!h]
    \centering
    \includegraphics[width=8cm]{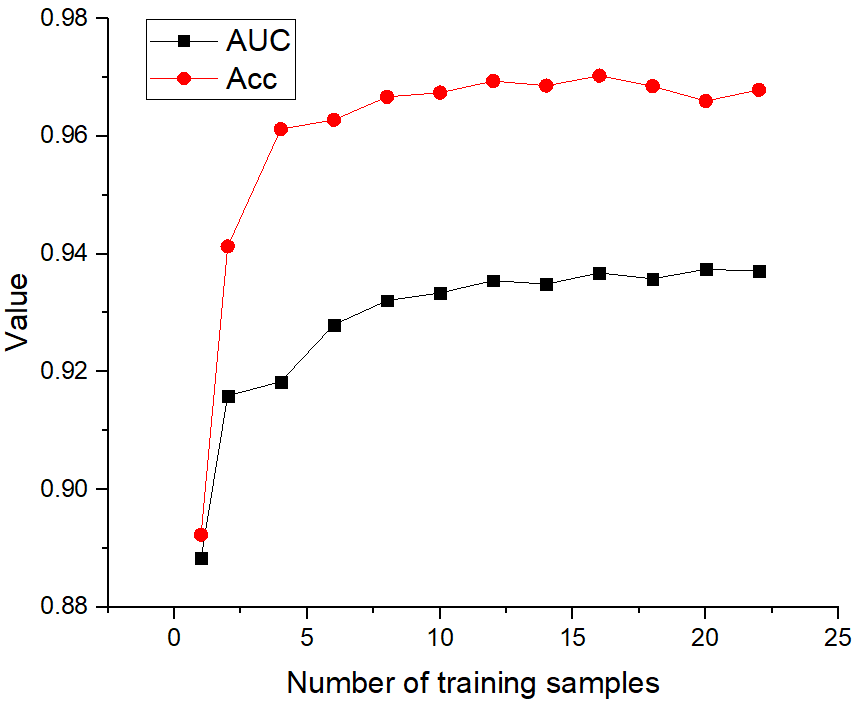}
    \caption{AUC and Acc values as functions of the number of training samples for the RPE shadow segmentation.}
    \label{unet}
\end{figure}
\subsubsection{Dataset} 
We employ the 30 OCT volumes from 30 subjects for the training and testing of the shadow localization and elimination pipeline. Each volume has $992\times512\times256$ voxels. They cover a field of view (FOV) of $6\times6$ mm$^2$ region and a imaging depth of around $2$ mm. We randomly divided these volumes into 25 testing sets and 5 evaluation sets. The choroid and RPE layers of these volumes were segmented with the Bio-Net. Then the \textit{en face} RPE projection images were manually annotated by two medical experts with pixel-level precision.
\subsubsection{Implementation} 
The shadow localization and elimination pipeline is also implemented using the PyTorch library \cite{paszke2019pytorch}. For training the U-Net in the shadow localization, we employ the Adam optimizer \cite{kingma2014adam} for training. The initial learning rate is set to 0.0001. Then we gradually decrease the learning rate with a momentum of 0.9. We further enhance the segmentation result of the U-Net with six iterations of dilation and erosion.\\
\indent For the shadow elimination, the Deshadow-Net is pre-trained with the neutral scene datasets in \cite{nazeri2019edgeconnect} then fine-tuned with our choroid datasets. The training of the model is divided into three stages: the edge model, the inpainting model, and the joint model, as suggested in the original implementation\footnote{https://github.com/knazeri/edge-connect}.
\subsubsection{Evaluation Metrics}
We employ the IOU, Acc, Sen and area under the curve (AUC) to evaluate the segmentation performance of the U-Net. Because no ground truth is available for the shadow elimination task, we employ the vessel density (VD) in the evaluation, which is an indirect but clinically useful metrics. We follow the calculation of the VD in \cite{jia2012quantitative} as 
\begin{equation}
VD = \frac{\int_A V \,dA}{\int_A \,dA},  
\end{equation}
Where $A$ is the region of interest (ROI). Here it refers to the $6\times6$ mm$^2$ centered on fovea. $V$ is the binarized vessel map. For an arbitrary pixel, if it belongs to a vessel, $V=1$, otherwise $V=0$. 
\begin{figure*}[!h]
    \centering
    \includegraphics[width=16cm]{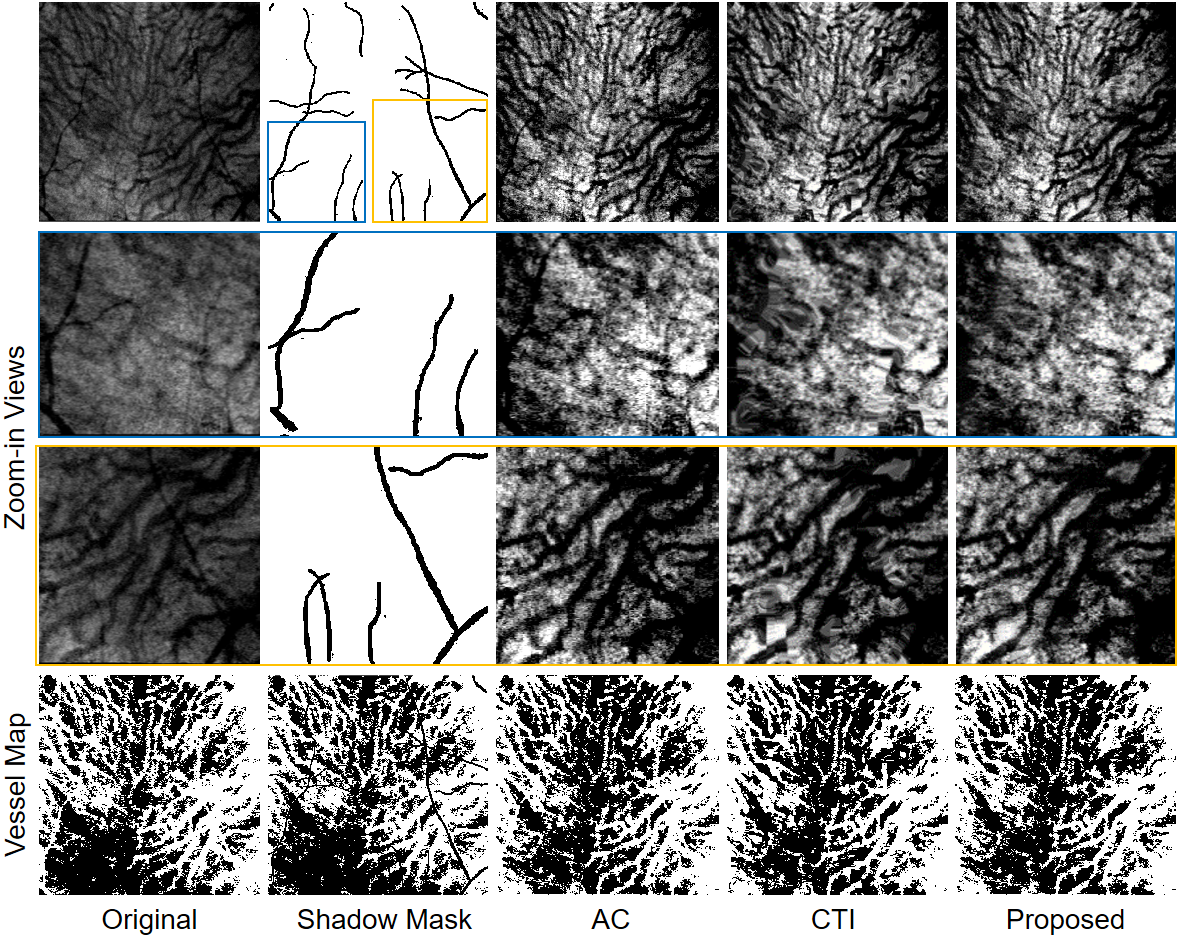}
    \caption{The visual examples of the shadow elimination. From left to right: the original \textit{en face} choroid image, the shadow mask, and the shadow elimination results using the AC \cite{girard2011shadow}, our previous work using the CTI \cite{yang2019enhancing}, and the proposed method in this work, respectively. The second and third rows are the zoom-in views inside the blue and yellow boxes in the first row, respectively. The last row are the corresponding vessel maps.}
    \label{shadow2}
\end{figure*}
\subsubsection{Results}
Figure~\ref{shadow1} demonstrates an example of using the U-Net in the segmentation of the retinal vessel shadows on the RPE projection image. The U-Net output is further processed to retrieve the shadow mask using dilation and erosion.  The U-Net achieves superior performance on the vessel shadow segmentation with a Acc of 0.969, a AUC of 0.938, a IoU of 0.901, and a Sen of 0.967. After the dilation and erosion, the vessel caliber is around 5 pixels wider than that of the original shadow, which makes sure the shadow could be completely removed by the Deshadow-Net.\\
\indent Ronneberger \textit{et al.} have demonstrated the U-Net was capable of achieving excellent segmentation performance with tens of training samples \cite{ronneberger2015u}. Here we found the U-Net could be efficient with less training samples. Figure~\ref{unet} demonstrates the AUC and Acc values as functions of the number of training samples for the RPE shadow segmentation. As shown in the figure, the AUC and Acc values are close to 0.9 with a single training sample. The two metrics trend to be stable when the number of training images is large than 5, which may be related to the high uniformity of the morphological patterns of the retinal vessel shadows among different OCT volumes.\\
\indent We compare the proposed shadow elimination pipeline with the A-line based AC algorithm \cite{girard2011shadow} and our previous implementation of this shadow localization and elimination pipeline, which used the CTI as the inpainting algorithm \cite{yang2019enhancing}. Figure~\ref{shadow2} demonstrates the visual examples of the shadow elimination. From left to right: the original \textit{en face} choroid image, the shadow mask, and the shadow elimination results using different methods. The second and third rows are the zoom-in views inside the blue and yellow boxes in the first row, respectively. The last row are the corresponding vessel maps.\\
\indent As shown in the figure, the original choroidal vasculature is conterminated by the retinal vessel shadows at the locations shown in the shadow mask. Inside the zoom-in views, the AC could enhance the contrast of the choroidal vessels and minimize small vessel shadows but could not get rid of the large vessel. Using the localization and elimination strategy, both the large and small vessel shadows could be thoroughly eliminated, but as shown in the zoom-in views, the CTI introduces unnatural artefacts compared with the proposed method.\\
\indent The vessel shadows could be treated as the real vessels in clinical assessment, which would cause the overestimation of the VD. The calculated VD values of the vessel maps in the last row of Fig.~\ref{shadow2} are: 0.510 for the original choroid, 0.504 for the AC, 0.501 for the CTI, and 0.500 for the proposed method. We also calculate a VD of 0.499 without including the shadow areas. The results are in accordance with the overestimation assumption, in which the original image has the highest VD. The AC method could eliminate part of the shadows thus lower the VD. The CTI and proposed method could further lower the VD because they remove the shadows completely. Besides, their VDs are very close to that of the masked vessel map, which indicate the effectiveness of this shadow localization and elimination pipeline. We checked the VDs of other testing datasets,which follow the exactly same trend. However, because the variation of the VDs among different eyes are much larger than that of the shadow elimination, we did not include their average values and standard deviations here. 
\begin{figure*}[!h]
    \centering
    \includegraphics[width=16cm]{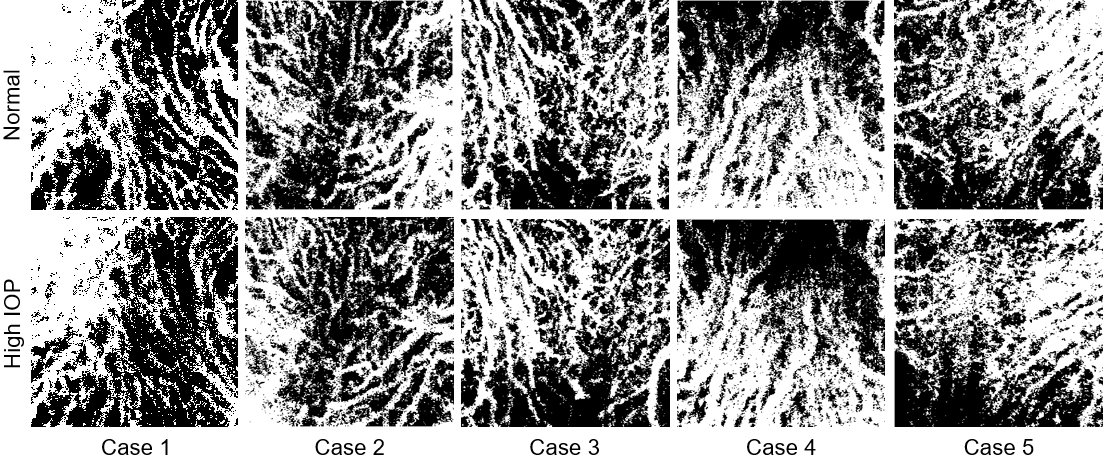}
    \caption{Demonstration of the vessel maps of 5 study cases in normal (top) and high IOP states (bottom). The reduction in blood flow could be observed in the high IOP state.}
    \label{iop}
\end{figure*}
\subsection{Application in a Clinical Prospective Study}
Primary angle-closure glaucoma (PACG) is prevailing not only in East Asia but also in overseas Chinese and Eskimos \cite{yip2006ethnic}. The patients with PACG were found to have higher IOP and thicker choroids than normal controls \cite{zhou2014increased}. Previous studies shown the changes of the choroid thickness and blood flow might be associated with the PACG \cite{chen2014changes}, but the initial mechanism underlying angle closure has not been fully understood. Here we applied the proposed method in a clinical prospective study, which quantitatively detected the changes of the choroidal in response to IOP elevation \cite{li2019upside}.
\subsubsection{Data Collection}
We recruited 34 healthy volunteers with the ages ranging from 18 to 30 years old, with no previous history of IOP exceeding 21 mm Hg. The participants were volunteers recruited mainly from the Zhongshan Ophthalmic Center at Sun Yat-sen University Medical School, and nearby communities in Guangzhou, China. The study was approved by the Ethical Review Committee of the Zhongshan Ophthalmic Center and was conducted in accordance with the Declaration of Helsinki for research involving human subjects. \\
\indent A swept-source OCT system with the A-line rate of 100 kHz (DRI OCT-1 Atlantis, Topcon, Japan) was employed to collect data from both of their eyes. We used the $6\times6$ mm$^2$ FOV volumetric scan protocol centered on fovea. Each of the volumes contains 256 B-scans and each of the B-scans has 512 A-lines. Each A-line contains 992 data points uniformly distributed in a depth range of $\sim3$ mm. \\
\indent To simulate the state of high IOP, after taking the baseline scans in a normal sitting position, each of the volunteers was asked to take scans in upside-down position. The average IOP was increased to $34.48\pm5.35$ mm Hg because of the upside-down, compared with the average IOP of $15.84\pm1.99$ mm Hg at the normal position. A total of 136 OCT volumes were acquired (34 volunteers, 68 eyes, normal and high IOP).
\subsubsection{Results}
With the proposed knowledge infused deep learning method, we processed this clinical dataset to retrieve the thickness of the choroid and the VD in normal and upside-down states. The choroid thickness is the average value of the $6\times6$ mm$^2$ FOV. We summarize their statical averages and standard deviations in Table \ref{q2}.\\
\begin{table}[!h]
\caption{Statics results of the clinical perspective study}\label{q2}
\centering
 \begin{threeparttable}
\begin{tabular}{llll}
\hline
 & IOP (mm Hg) & CT\tnote{*} ($\mu$m) & VD\\
 \hline
 Normal &  $15.84\pm1.99$ &  $226.39\pm52.44$ &  $0.511\pm0.171$\\
 Upside-Down & $34.48\pm5.35$ & $238.34\pm54.84$ & $0.488\pm0.164$\\
 \hline
\end{tabular}
 \begin{tablenotes}
        \footnotesize
        \item[*] CT: choroid thickness.
      \end{tablenotes}
    \end{threeparttable}
\end{table}
\begin{figure*}[!h]
    \centering
    \includegraphics[width=13cm]{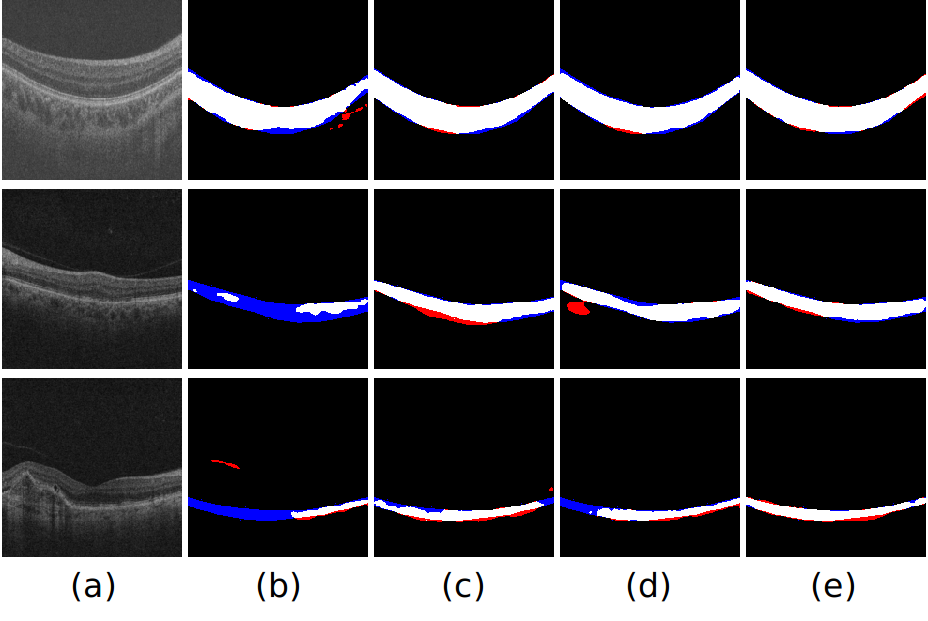}
    \caption{Ablation study of the Bio-Net. From left to right: (a) is the input OCT B-scans, and from (b) to (e) are the segmentation results using the baseline method (U-Net), baseline+the biomarker prediction (Bio) module, baseline+the global multi-layer segmentation (GMS) module, and the  proposed Bio-Net (baseline+Bio+GMS), respectively. White indicates the choroid region that is correctly segmented, red indicates excessive segmentation, and blue indicates insufficient segmentation.}
    \label{choroid2}
\end{figure*}
\indent As the IOP increases from $15.84\pm1.99$ mm Hg during the normal sitting state to $34.48\pm5.35$ mm Hg during the upside-down, the choroid becomes thicker while the VD decreases. Both of them has a p value of $<0.001$. These results provide evidence about the relationship between choroid expansion and shallowing of the anterior chamber, which may be of relevance for the pathogenesis of the PACG.\\
\indent Figure~\ref{iop} demonstrates the vessel maps of 5 study cases from the clinical dataset in normal (top) and high IOP states (bottom). In OCT imaging, the change on the vessel map is related to the change in blood flow. On these vessel maps, we could observe the reduction in blood flow in the high IOP state.
\section{Discussions}
We further analyze and discuss the details of the proposed method in this section, including the ablation study of the Bio-Net, the comparison of different inpainting methods, and the limitations of this work.
\subsection{Ablation Study of Bio-Net}
\begin{figure*}[!h]
    \centering
    \includegraphics[width=17cm]{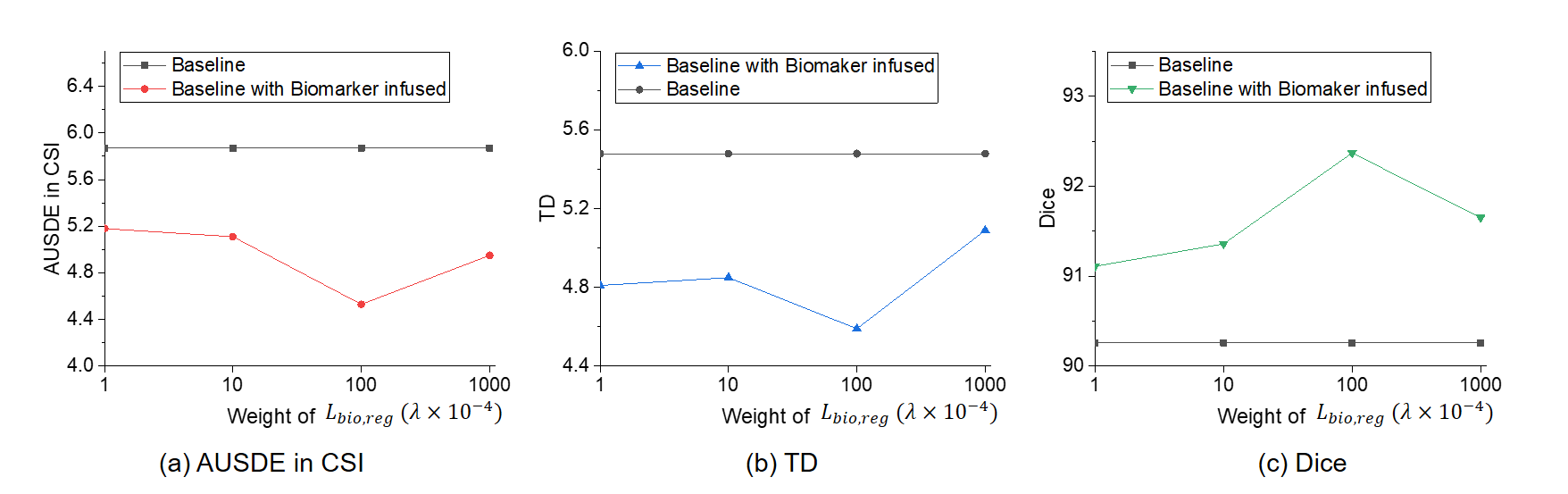}
    \caption{Effectiveness of the proposed Bio module by changing its hyper-parameter spanning four orders of magnitude. (a), (b), and (c) are the results using AUSDE in CSI, TD, and Dice as measures, respectively.}
    \label{Bio}
\end{figure*}

\begin{figure*}[!h]
    \centering
    \includegraphics[width=17cm]{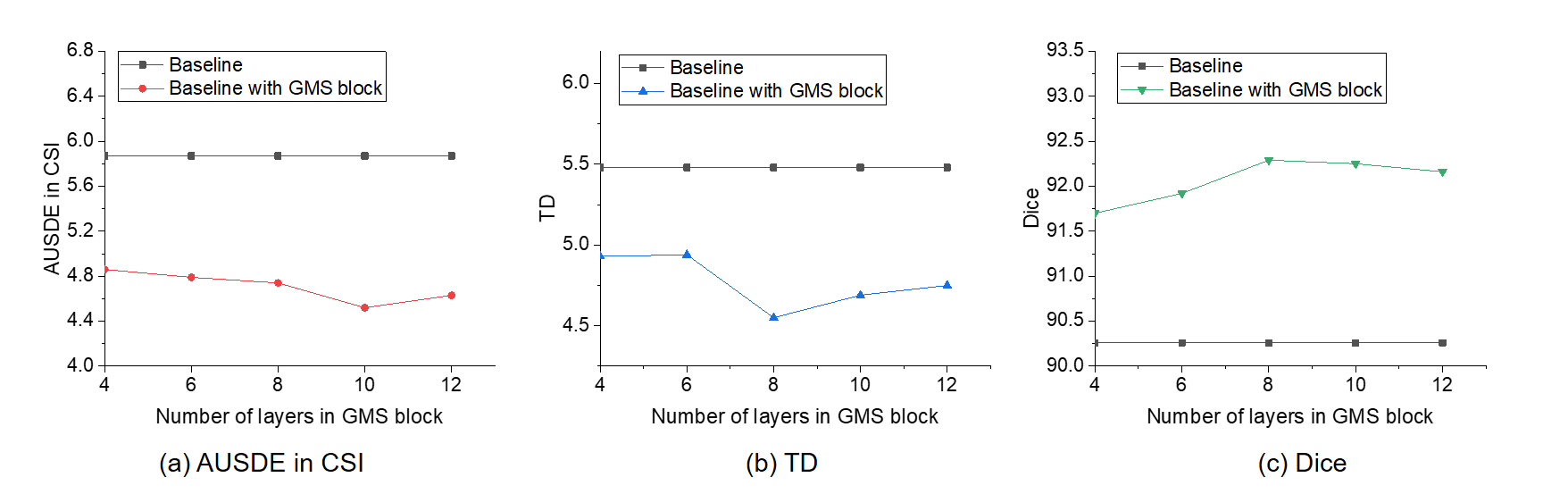}
    \caption{Effectiveness of the proposed GMS module by changing the number of the segmented layers from 4 to 12. (a), (b), and (c) are the results using AUSDE in CSI, TD, and Dice as measures, respectively.}
    \label{GMS}
\end{figure*}

\begin{table*}[!h]
 \caption{Quantitative comparison of different inpainting algorithms}\label{q4}
 \centering
 \begin{tabular}{lllll}
 \hline
 & Masked & EBI \cite{criminisi2004region}  & CTI \cite{bornemann2007fast} & GAN \cite{nazeri2019edgeconnect} \\
\hline
SSIM & $0.778\pm 0.085$ & $0.901\pm 0.043$ & $0.931\pm 0.015$ & $0.946\pm0.013$\\
PSNR & $11.851\pm2.605$ & $25.988\pm3.117$ & $29.352\pm3.712$  &$30.615\pm3.824$ \\
MSE ($\times10^3$) & $3.464\pm1.895$ & $0.165\pm0.092$ & $0.109\pm0.082$&$0.091\pm0.071$ \\ 
 \hline
 \end{tabular}
 \end{table*}
To evaluate the effectiveness of the global multi-layers segmentation module and the biomarker prediction Net, we combine them with the U-Net respectively.\\
\indent Fig. \ref{choroid2}, Fig. \ref{Bio} and Fig. \ref{GMS} illustrate the effectiveness of the biomarker prediction network and the global multi-layers segmentation module. In the experiment, we take the U-Net as a baseline, the table demonstrates that the infusion of the biomarker prediction network can lead to an improvement on the choroid segmentation task, as the DI increases from $88.36\%$ to $90.27\%$ and the AUSDE decreases from $8.01$ pixels to $6.46$ pixels. On the other hand, the performance of the U-Net added by the global multi-layers segmentation module makes the IOU increases from $79.14\%$ to $81.34\%$ and AUSDE decreases from $8.01$ pixels to $6.54$ pixels. Meanwhile, the AUSDE is $3.50$ pixels lower and Sen $1.79\%$ higher than only GMS module employed, which demonstrates that the global-to-local network works better than a single global multi-layers segmentation module or a single local choroid segmentation module. \\
\indent We also examined the effectiveness of the proposed Bio module by changing its hyper-parameter spanning four orders of magnitude as show in Fig.~\ref{Bio}. For different metrics including AUSDE, TD, and Dice, the best performance is all achieved at a weight of $\mathcal{L}_{bio,reg}$ of 0.01. The results of the Bio-Net are always better than those of the U-Net baseline. For the GMS module, we checked the influence of the segmentation layers from 4 to 12 layers as shown in Fig.~\ref{GMS}. The observation of the results is very similar to that of Fig.~\ref{Bio}. Through these ablation studies, we may be confident about the proposed biomarker-infused global-to-local segmentation network.

\subsection{Quantitative Comparison of inpainting Methods}
In this work, we employ the GAN inpainting method \cite{nazeri2019edgeconnect} for the shadow elimination task. We have compared it with our previous implementation \cite{yang2019enhancing} using the CTI \cite{bornemann2007fast} as shown in Fig.~\ref{shadow2}. It shows both of the inpainting methods could completely eliminated the shadows and fill the shadow areas with neutral extensions of surrounding contents. We also could notice the CTI create some locate artefacts. Here we further compare the performance of these inpainting method quantitatively. We also include the EBI \cite{criminisi2004region} in the comparison.\\
\indent Due to the absent of the ground-truth choroidal vasculature, the quantitative comparison of the inpainting methods can not directly implemented. Thus we created artificial retinal vessel mask with the vessel widths slightly wider than the real shadows. The artificial mask combining with the repaired choroid images were used as the input to evaluate the inpainting algorithms. The widely-used image similarity measures including structure similarity index (SSIM), peak signal to noise ratio (PSNR), and mean squared error (MSE) \cite{cheng2016speckle} are employed as quantitative metrics. We used the masked images as the baseline. \\
\indent As shown in Table~\ref{q4}, the GAN inpainting method outperforms the CTI and EBI methods on all the metrics. However, its advantages are marginal. Because all of the inpainting methods achieve SSIMs $> 0.9$, which suggest that all of them could be suitable choice for the proposed localization and elimination pipeline. 
\subsection{Limitations of This Work}
The methodology proposed in this work was developed under the requirement of quantifying subtle changes of the choroid blood flow in response to the IOP elevation, as described in Section \uppercase\expandafter{\romannumeral4}-C. Because it is a clinical perspective study, only healthy volunteers were involved. So we did not take the pathological changes such as CNV into consideration. Even though, we found the existing methods were still need to be improved in the automatic choroid segmentation and vessel shadow removal, as discussed in Section \uppercase\expandafter{\romannumeral1} and \uppercase\expandafter{\romannumeral2}. Based on the observation of the clinical IOP dataset, we propose the Bio-Net for the automatic choroid segmentation and the shadow removal pipeline for quantifying choroidal vasculature. The usage of the proposed method leaded to a significant clinical finding that a high IOP would cause the thicken of the choroid layer and the decrease of choroidal blood flow, which may help us to understand the mechanism of glaucoma \cite{li2019upside}. To examine the performance of the Bio-Net in the presence of the pathological changes of choroid, we included CNV cases in the training and testing and found our proposed method still outperformed the state-of-the-arts as demonstrated in Fig.~\ref{choroid1} and Table~\ref{q1}. However, this verification is not enough since there are several other types of the pathological changes of choroid, such as DR and pathological myopia. Besides, a very limited sample size was collected at the time of submission. We need to keep collecting more cases with better diversity. Additionally, variance quantification of same subject over time would be extremely useful for delicate clinical applications.\\
\indent For the shadow removal pipeline, a major limitation is the method of determining the positions of vessel shadows. For the cases collected in the clinical perspective study \cite{li2019upside}, the RPE layer is undamaged, so the shadow localization is straightforward. The intuition behind our method also brings new perspective to this vessel shadow removal task, as discussed in Section \uppercase\expandafter{\romannumeral1} and \uppercase\expandafter{\romannumeral2}. However, if the RPE layer is damaged due to CNV or other types of ocular diseases, our method will be invalid. A neutral thought is to use the layers above the RPE, such as ONL, to localize the shadows. But we found it might not be feasible because the correspondence between the shadows in retinal and choroidal layers was broken due to the strong light absorption at the lesion areas. So it may be necessary to further combine the A-line-based method and our method or develop specific methods, for handling the RPE-damaged cases.\\
\indent Another major limitation of our method is its possible failure in tackling the OCT B-scans including optic nerve head (ONH). As stated above, this work was inspired by the clinical perspective study in \cite{li2019upside}. Thus we only have the macular-centered OCT B-scans and labels. We cannot assure that the Bio-Net and vessel removal pipeline still perform well in the presence of the ONH, which has been listed in our future works. The capability of segmenting both the macular and ONH will enable the generalization of segmenting wide-field OCT scans. Because the early signs of many ocular diseases such as DR emerge at the edge of the retina. This technique will be useful for the screening of ocular diseases among a large population.

\section{Conclusion}
In this paper, we have developed an automatic method for the segmentation and visualization of the choroid, which combined deep learning networks with prior medical and OCT imaging knowledge. We have proposed the Bio-Net for the choroid segmentation, which is a biomarker infused global-to-local network. It outperforms the state-of-the-art choroid segmentation methods. For eliminating the retinal vessel shadows, we have proposed a deep learning pipeline, which firstly locates the shadows using anatomical and OCT imaging knowledge, then removes the shadow using a two-stage GAN inpainting architecture. Compared with the existing methods, the proposed method has superiority in shadow elimination and morphology preservation. We have further applied the proposed method in a clinical prospective study, which quantitatively detected the changes of the choroid in response to IOP elevation. The results show it is able to detect the structure and vascular changes of the choroid efficiently.
\section*{Acknowledgement}
We would like to thank all the reviewers for their insightful comments and concerns which stimulate us to improve our work.
\bibliographystyle{IEEEtran}
\bibliography{ref2}{}
%
%
%
%
%
%
\end{document}